\documentclass[letterpaper, 10 pt, conference]{ieeeconf}  

\IEEEoverridecommandlockouts                              
\usepackage{amsmath,graphicx}
\let\labelindent\relax

\usepackage{enumitem}
\usepackage{siunitx}
\usepackage{verbatim}
\title{\LARGE \bf
Tissue Classification During Needle Insertion Using Self-Supervised Contrastive Learning and Optical Coherence Tomography
}

\author{Debayan Bhattacharya$^{1,2}$, Sarah Latus$^{1}$, Finn Behrendt$^{1}$, Florin Thimm$^{1}$, Dennis Eggert$^{2}$, \\ Christian Betz$^{2}$,  
Alexander Schlaefer$^{1}$ 
\thanks{$^{1}$Institute of Medical Technology and Intelligent Systems, Hamburg University of Technology, Hamburg, Germany
        {\tt\small debayan.bhattacharya@tuhh.de}}%
\thanks{$^{2}$Clinic for Ears, Nose and Throat, University Medical Center Hamburg-Eppendorf, Hamburg, Germany}%
}

\begin{document}

\maketitle
\thispagestyle{empty}
\pagestyle{empty}

\begin{abstract}

Needle positioning is essential for various medical applications such as epidural anaesthesia. Physicians rely on their instincts while navigating the needle in epidural spaces. Thereby, identifying the tissue structures may be helpful to the physician as they can provide additional feedback in the needle insertion process. To this end, we propose a deep neural network that classifies the tissues from the phase and intensity data of complex OCT signals acquired at the needle tip. We investigate the performance of the deep neural network in a limited labelled dataset scenario and propose a novel contrastive pretraining strategy that learns invariant representation for phase and intensity data. We show that with 10\% of the training set, our proposed pretraining strategy helps the model achieve an F1 score of 0.84±0.10 whereas the model achieves an F1 score of 0.60±0.07 without it. Further, we analyse the importance of phase and intensity individually towards tissue classification.

\end{abstract}


\section{Introduction}
\label{sec:intro}
Precise positioning of needles is essential for various medical applications. For example, during epidural anesthesia the physician needs to navigate a needle to the epidural space, which has a width of a few millimeter. Hence, identification of the tissue structures passed during insertion is important to estimate the position of the needle. Image-guided approaches, e.g., using ultrasound (US)~\cite{Pesteie2018USEpidural}, as well as electromagnetic or optical needle pose tracking have been investigated to assist needle placement.
To address inaccuracies of these external needle navigation methods~\cite{McLeod.2021}, approaches to estimate tissue properties from the needle tip have been studied, e.g. by sensing the forces acting at the needle tip~\cite{Beisenova2018FBG,Gessert2019ForceNeedle} or integrating miniaturized imaging probes~\cite{Wang2022Epidual}. 
Recently, the application of optical coherence tomography (OCT), an imaging modality with high spatial and temporal resolution, has been proposed. It enables depth-resolved scanning of tissue structures by measuring the reflections of infrared light. The acquired complex OCT signal can be split into intensity and phase data, with the latter resolving tissue motion in sub-pixel range. However, imaging depth of OCT is limited to a few millimeters and interpretation of image data requires time consuming analysis by experienced clinician who may not be available every time. 

Hence, supervised deep learning methods have been proposed that use OCT intensity data for tissue classification~\cite{Wang2022Epidual,Otte2013LSTMAscanClass}
and boundary tracking~\cite{Guo2022CNNBoundaryTracking} during needle insertion. In addition, deep learning based rupture detection~\cite{Latus2021Rupture} has been proposed, combining intensity and phase data. Still, manual labeling of acquired OCT datasets is time consuming and requires additional information on the needle pose and surrounding tissue structures. These ground-truth information are accessible for phantom studies with optimized needle tracking conditions. But gaining ground-truth information during in-vivo needle insertions is complicated and hampered by decreased imaging resolutions. This motivates the investigation of self-supervised learning strategies for tissue classification during needle insertion so as to reduce the labelling effort. 


Self-supervised learning has been proposed to alleviate the problem of time-consuming labelling by learning visual features from unlabelled dataset  \cite{DBLP:journals/corr/ZhangIE16,DBLP:journals/corr/NorooziF16,DBLP:journals/corr/PathakKDDE16,Kumar2022}. A popular self-supervised learning method SimCLR \cite{https://doi.org/10.48550/arxiv.2002.05709} learns invariant representation of an image and its augmented version. Using the InfoNCE loss \cite{DBLP:journals/corr/abs-1807-03748}, the network learns invariant representations for the image and its augmented version. In our case, we have intensity and phase data from the complex OCT signal. Augmentations such as rotation and random crop may not be ideal for our data as the augmented views will represent signals that are unrealistic to generate by needle insertions. As such, meaningful representations will not be learned in the pretraining step. Instead, to learn better representations, we align phase and intensity features within a time window. 
Empirically, we demonstrate that this pretraining strategy is helpful for tissue classification especially in limited labelled dataset scenario.  

In conclusion, our contributions are three fold.
First, we propose a deep learning algorithm which classifies tissue class by jointly using phase and intensity data. Second, to circumvent the issues arising from limited labelled dataset, we propose a novel self-supervised contrastive pretraining strategy that increases the classification performance in an extremely low labelled dataset scenario. Third, analogous to the rupture detection study~\cite{Latus2021Rupture}, we investigate the importance of the phase and intensity data on the classification performance by evaluating only using phase or intensity data.

\section{Materials and Methods}
\label{sec:method}
\subsection{Data acquisition and preparation}
\label{ssec:dataAcq}

\begin{figure}
    \centering
    \includegraphics[width=1.0\linewidth]{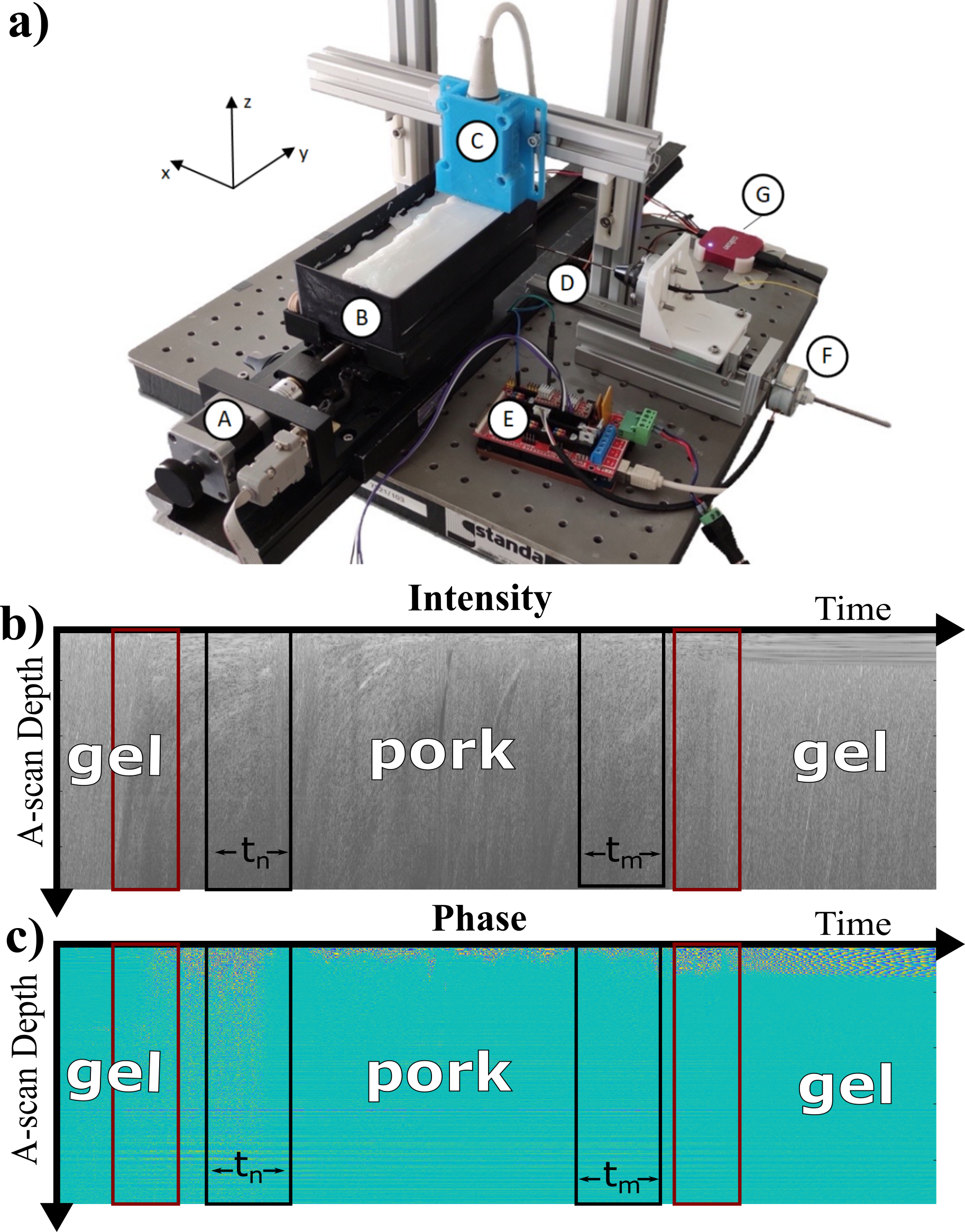}
    \caption{a) Experimental setup for data acquisition. We use a microcontroller (E) with attached stepper motor (A) to position soft tissue phantoms (B) relative to a fixed US probe (C). We insert our OCT needle (D) with a second stepper motor (F). A logic analyzer (G) records the trigger signals of all components. Visualisation of OCT b) intensity and c) phase M-scan with highlighted uncertainty windows (red).}
    \label{fig:setup}
\end{figure}
We design an experimental setup to perform reproducible needle insertions in soft tissue phantoms (Fig.~\ref{fig:setup}). We embed a cleaved optical fiber in an epidural Touhy needle to acquire one-dimensional scans (A-scans) at the needle tip with our OCT system (Telesto I, Thorlabs). To track the needle motion relative to soft tissue structures we additionally apply US imaging with a linear probe (Cicada, Cephasonics). We use two stepper motors to control the needle and phantom motion and conduct multiple needle insertions per phantom. We insert the needle with a velocity of \SI{1}{\milli\meter\per\second} and simultaneously acquire OCT A-scans and US images with frequencies of \SI{91}{\kilo\hertz} and \SI{60}{\hertz}, respectively. For data synchronization we record the trigger signals of the motor driver, OCT and US system.
Our soft tissue phantoms are composed of beef, pork, or turkey meat embedded in gelatin of \SI{5}{\percent} density to imitate tissue structures punctured during epidural punctures similar to~\cite{Beisenova2018FBG}.
We perform three data preprocessing steps. 
First, we display the A-scans sequentially recorded during insertion side by side over time to obtain M-scans, 
considering either OCT intensity or phase data (Fig.~1 (b) and (c)). Second, we average the M-scans over time with a window size of 1000 A-scans and apply spatial cropping. 
Third, we extract the needle position relative to the tissue structures from the US images to label the OCT M-scans. 
To consider inaccuracies for transitions between tissue classes we define an uncertainty window to be excluded from the labelled dataset.

\subsection{Overall Framework}

Fig.~2 illustrates our proposed method. 
 Our proposed method is a deep neural network with two branches (\(f_{\theta }\), \(g_{\theta }\)), that processes the intensity and phase data respectively. We split the method in two parts: (i) Intensity and phase alignment using contrastive pretraining, (ii)  Classification task using \(c_{\theta }\) as the classification head.  
 \begin{figure}[htb!]
\centering
\begin{minipage}[b]{1\linewidth}
  \centering
  \centerline{\includegraphics[width=1.0\linewidth]{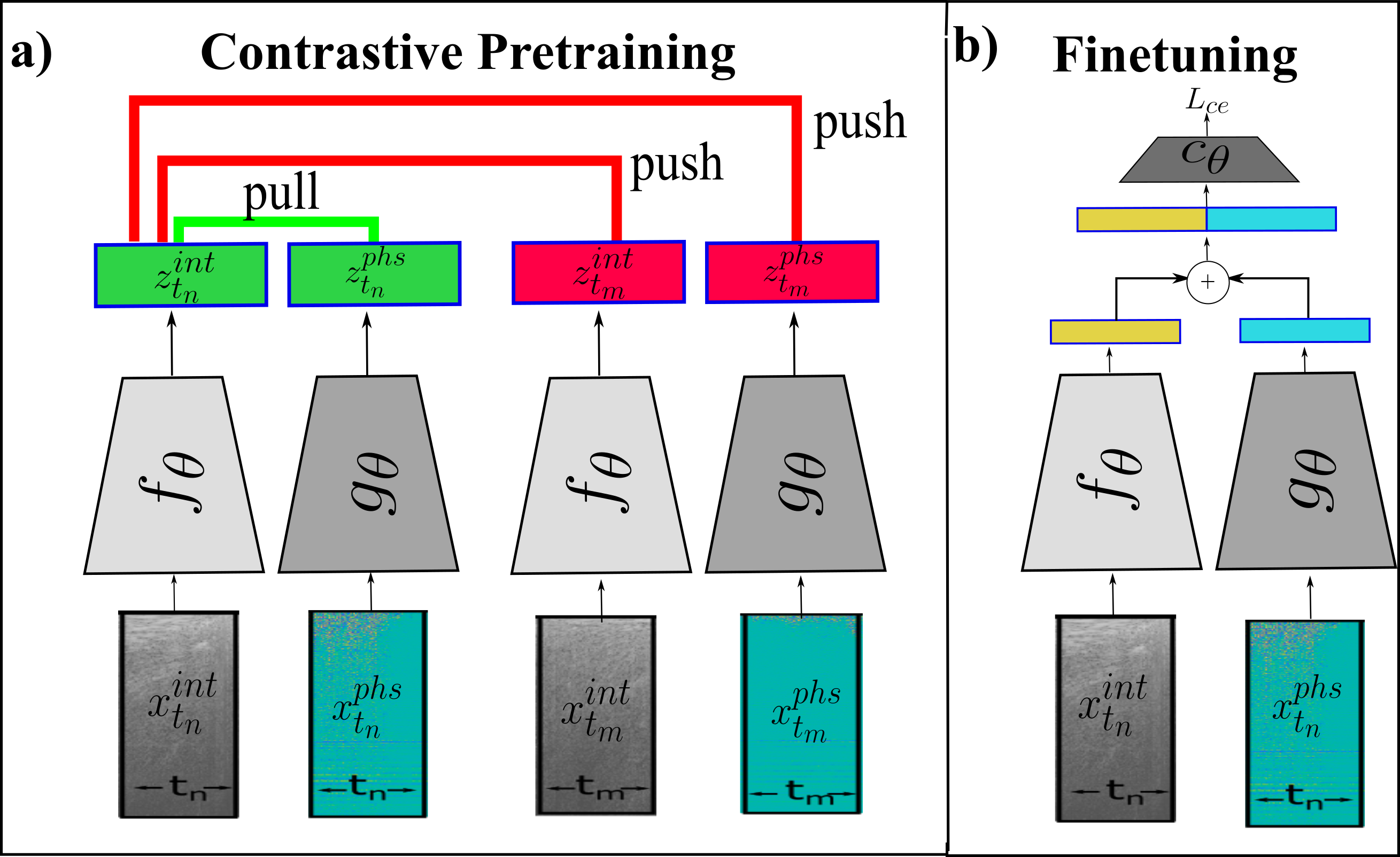}}
  \caption{(a) Our contrastive pretraining strategy  (b) Finetuning for tissue classification task.}\medskip
\end{minipage}
\label{fig:method}
\end{figure}

\textbf{(i) Intensity and phase alignment using contrastive pretraining:} Our positive pair consists of intensity and phase data in the same time window \(t_{n}\). Formally, let \(x^{phs}_{t} \in R^{H \times W \times 3}\) and \(x^{int}_{t} \in R^{H \times W \times 3}\) be phase and intensity crops at time window \(t\) respectively. It is to be noted that the phase and intensity crops are originally single channel images. The information is copied to the 2 channels to make it a 3 channel input for \(f_{\theta}\) and \(g_{\theta}\). We pass \(x^{int}_{t}\) through \(f_{\theta}\) and  \(x^{phs}_{t}\) through \(g_{\theta}\). This results in phase and intensity feature vectors \(z\) in hyperspace \(R^{D}\), with \(z^{int}_{t} = f_{\theta}(x^{int}_{t}) \) and \(z^{phs}_{t} = f_{\theta}(x^{phs}_{t})\) respectively. \(sim\) computes the cosine similarity between two vectors. We train the models using the following contrastive loss: 

\begin{equation}
\begin{split}
  & \mathrm{L_{\mathrm{c}}(i)} = - \mathrm{log} \frac{\mathrm{e}^{\mathrm{sim(z^{int}_{i},z^{phs}_{i})}/\tau}}{\mathrm{e}^{\mathrm{sim(z^{int}_{i},z^{phs}_{i})}/\tau} + S_1 +  S_2} \hspace{0.5cm}\text{with} \\
  &S_1 = \sum\limits_{j \neq i} \mathrm{e}^{\mathrm{sim(z^{int}_{i},z^{phs}_{j})}/\tau}, S_2 = \sum\limits_{j \neq i} \mathrm{e}^{\mathrm{sim(z^{int}_{i},z^{int}_{j})}/\tau}.
\end{split}
\end{equation}

\textbf{(ii) Classification Task:} 
For our experiments, we consider four tissue classes: gelatin, pork, beef, turkey. Our mini-batch consists of randomly sampled intensity and phase data crops belonging to a time window \(t_{n}\) from multiple phantom insertions. We train \(c_{\theta}\), \(f_{\theta}\) and \(g_{\theta}\) using cross-entropy loss.

\section{Experiments}
\label{sec:experiments}

We test the feasibility of our proposed method using our in-house dataset. We compare our method against models trained using randomly initialised weights ("Scratch") and using ImageNet \cite{deng2009imagenet} weights in different training set sizes. We compare all the methods with 10\%, 20\%, 30\%, 60\%, 80\% and 100\% of the training set. Further, we compare the importance of phase and intensity features towards tissue classification by solely using \(f_{\theta}\) or \(g_{\theta}\). 
 
 \textbf{Dataset split: } We have 34, 14 and 18 needle insertions on soft-tissue phantom with beef, pork and turkey respectively. We split the needle insertions into 80:10:10 train, validation and test split stratified on tissue class. From each M-scan, we crop intensity and phase data by passing a rectangular window of width 256 and height 250 along the time dimension of the M-scans. Examples of the extracted phase and intensity data are shown in Fig~1 (c) and (d). Note that the rectangular window does not overlap in area with its last position as it slides across the time dimension. In total, 269, 169, 64 and 115 phase and intensity crops of gelatin, beef, pork and turkey are extracted. For our contrastive pretraining, we use 100\% of the training set without labels. 
 
\textbf{Implementation Detail:} \(f_{\theta }\) and \(g_{\theta }\) are two identical ResNet18 \cite{DBLP:journals/corr/HeZRS15} models. \(c_{\theta }\) is a multilayer perceptron with 3 layers having dimension 1024, 512 and 4 with a ReLu in between when using both \(f_{\theta }\) and \(g_{\theta }\) and 512, 512 and 4 when using \(f_{\theta }\) or \(g_{\theta }\). We train our network using Adam optimizer \cite{DBLP:journals/corr/KingmaB14}. We use a batch size of 28. D is set to 512 in \(R^{D}\) and \(\tau = 0.1\). 
We train all our models for 100 epochs. Our reported metrics are weighted Average Precision (AP) and F1. All our experiments are trained using 3-fold cross validation and we report the mean and standard deviation of the metrics.  

\section{Results}
\label{sec:results}

The results of our experiments are reported in Tab.~\ref{tab:result}. We report classification performance in 10\%, 20\% and 30\% of the training set. First, AP and F1 scores are best for models trained using our pretraining strategy in all the limited dataset scenarios. 

\begin{table}[htbp]
\caption{Results in limited labelled dataset scenario}
\label{tab:result}
\begin{center}
\resizebox{0.85\linewidth}{!}{%
\begin{tabular}{ |c|c|c|c|c| } 

\hline
\% Training Set & Method & AP & F1 \\
\hline
10 & Scratch & 0.73±0.07 & 0.60±0.07 \\ 
& ImageNet & 0.85±0.10 & 0.76±0.1 \\ 
& Pretrained & \textbf{0.94±0.05} & \textbf{0.84±0.10} \\ 
\hline
20 & Scratch & 0.88±0.04 & 0.69±0.05 \\ 
& ImageNet & 0.95±0.03 & 0.84±0.04 \\ 
& Pretrained & \textbf{0.97±0.01} & \textbf{0.89±0.03} \\ 
\hline
30 & Scratch & 0.93±0.05 & 0.84±0.05 \\ 
& ImageNet & 0.97±0.02 & 0.92±0.03 \\ 
& Pretrained & \textbf{0.99±0.006} & \textbf{0.95±0.02} \\ 
\hline

\end{tabular}}
\end{center}
\end{table}

\begin{figure}[htb!]

\begin{minipage}[b]{1.0\linewidth}
  \centering
  \centerline{\includegraphics[width=1.0\linewidth]{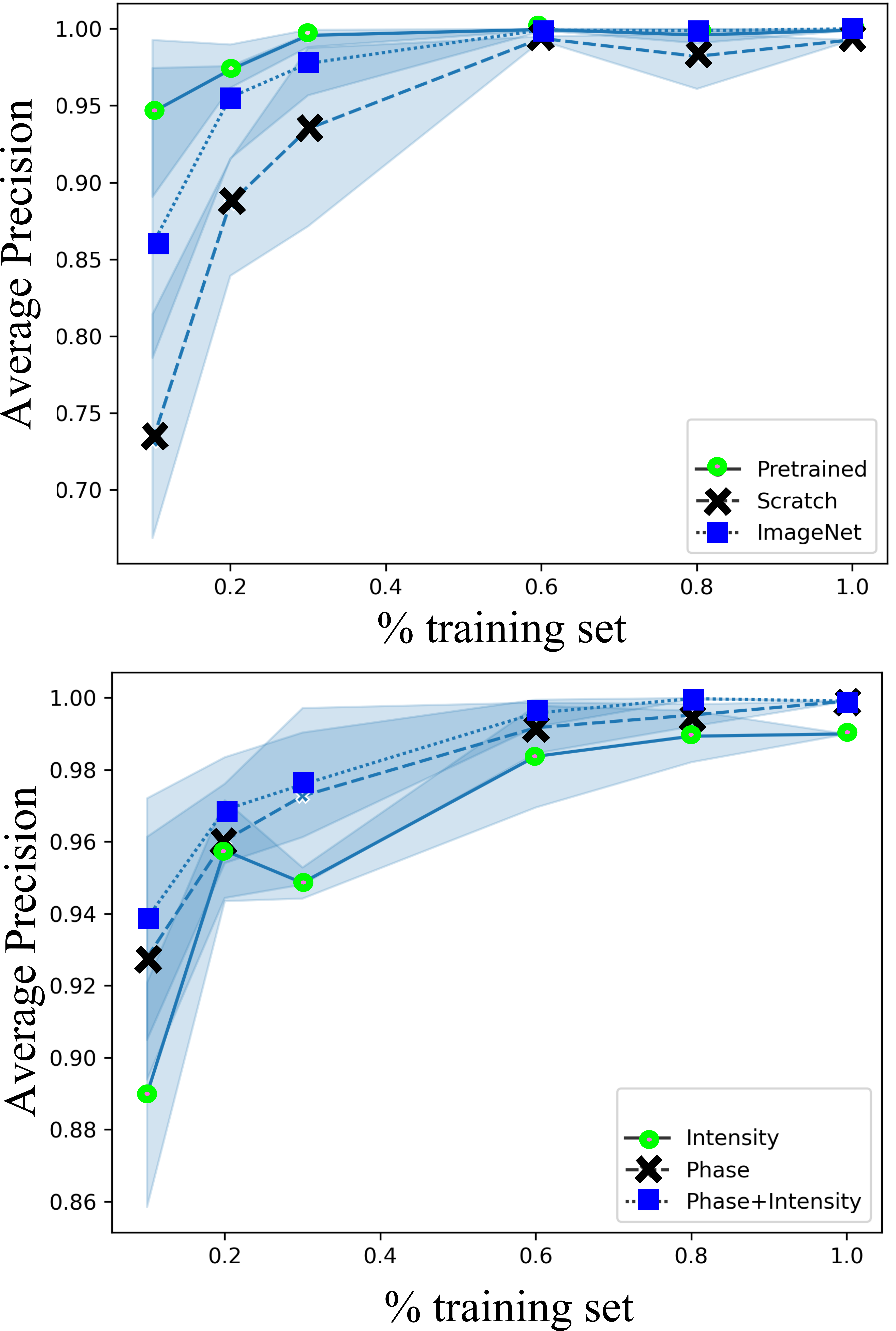}}
  \caption{(Top) Comparing Average Precision vs \% of training set  for models trained from scratch, with ImageNet weights and using our pretraining strategy  (Bottom) Average Precision vs \% of training set for models trained using only phase crop, only intensity crop and using both phase and intensity crop }\medskip
\end{minipage}
\label{fig:res}
\end{figure}

Fig.~3 (Top) shows that with infusion of data into the training stage, the three methods converge in performance.  Fig.~3 (Bottom) shows the relative importance of phase and intensity data towards classification. Combination of phase and intensity shows consistently highest tissue classification performance throughout. We also observe that phase data are more beneficial towards intensity. 


\section{Discussion and Conclusion}
\label{sec:discuss}

From the results in Tab.~\ref{tab:result} we observe that pretraining using our proposed strategy is the most beneficial for tissue classification in limited dataset scenario. 
We conjecture that the improvement in performance is because the model learns invariant representations for phase and intensity data and in doing so, the weights of the network are tuned to our target OCT dataset. The lowest performances are observed for the models trained from scratch. This can be attributed to overfitting of the models to the small labelled dataset. With respect to the models trained using ImageNet weights, we hypothesize that due to the large differences between ImageNet dataset and our OCT dataset, the ImageNet weights do not transfer well to our OCT dataset. 
Considering Fig.~3 (right), the differences in performance when training with intensity or phase data might be related to the fact that the speckle patterns in the intensity data are not necessarily clearly distinguishable when the tissue structures are compressed in front of the needle. In contrast the phase data, capable for sub-pixel motion detection, might contain tissue dependent features during compression.
Finally, from Fig.~3 we also conclude that with infusion of data, all the methods achieve competitive performance. 

In conclusion, we propose a deep learning model to classify tissue from complex OCT data. We show that combination of intensity and phase data is most beneficial and leads to the highest average precision. Further, we propose a novel self-supervised contrastive pretraining strategy that proves to be beneficial in low labelled dataset scenarios. We believe that this pretraining stategy may prove to be helpful in improving the classification performance for in-vivo needle insertion. 

\section{Compliance with ethical standards}
\label{sec:ethics}

No experiments were performed on humans or animals. No ethics approval was required. 



\section{Acknowledgments}
\label{sec:acknowledgments}

The authors have no conflicts of interests to report. This work has not been submitted for publication anywhere else. This work is funded partially by the i$^3$ initiative of the Hamburg University of Technology. The authors also acknowledge the partial funding by the Free and Hanseatic City of Hamburg (Interdisciplinary Graduate School) from University Medical Center Hamburg-Eppendorf.


\bibliographystyle{IEEEbib}
\bibliography{refs}

\end{document}